\documentclass[12pt]{article}
\usepackage[utf8]{inputenc}
\usepackage{amsmath}
\usepackage{amssymb}
\usepackage{url}
\usepackage{mathtools}
\usepackage{bbm}
\usepackage[margin=1.0in]{geometry}
\usepackage{natbib}
\usepackage{xcolor}
\usepackage{algorithm}
\usepackage{algpseudocode}
\usepackage{makecell}
\usepackage{hyperref}

\newcommand{\blinded}[1]{{\sc blinded for review}}

\title{Tractable Algorithms for Changepoint Detection in Player Performance Metrics}
\author{Amanda Glazer\thanks{Department of Statistics and Data Sciences, The University of Texas at Austin. Email: \texttt{amanda.glazer@austin.utexas.edu}}}
\date{}

\begin{document}

\maketitle

\begin{abstract}
We present a tractable framework for detecting changes in performance metrics and apply these methods to Major League Baseball (MLB) batting and pitching data from the 2023 and 2024 seasons. 
We propose a changepoint detection algorithm that combines a likelihood-based approach with split-sample inference to better control false positives, using either nonparametric tests or tests appropriate to the underlying data distribution.
These tests incorporate a shift parameter, allowing users to specify the minimum magnitude of change to detect.
We demonstrate the utility of this approach across simulation studies and several baseball applications: detecting changes in batter plate discipline metrics (e.g., chase and whiff rate), identifying velocity changes in pitcher fastballs, and validating velocity changepoints against a curated quasi-ground-truth dataset of pitchers who transitioned from relief to starting roles. Our method flags meaningful changes in 91\% of these `ground-truth' cases and reveals that, for some metrics, more than 60\% of detected changes occur in-season. While developed for baseball, the proposed framework is broadly applicable to any setting involving monitoring of individual performance over time.
\end{abstract}


\section{Introduction}
Aaron Judge, an outfielder for the New York Yankees, led Major League Baseball (MLB) in on-base percentage (OBP) during the 2024 season, finishing with an impressive OBP of .458 compared to the league average of .312. Early in the season, however, his OBP after his first 24 plate appearances was just .208, substantially lower than his end-of-season OBP. Intuitively, this discrepancy highlights the well-known issue that performance metrics based on small sample sizes can vary substantially, and that this variability typically decreases as sample size increases. This example naturally raises a fundamental question relevant not just in sports analytics but in any field that tracks performance over time: how can we efficiently detect statistically meaningful shifts in an observed performance metric, or determine when apparent changes are simply due to small sample variability?

Across sports, sample size in relation to metric stabilization and changepoints is a widely discussed topic with researchers often asking how large a sample is needed to be reasonably confident in the corresponding results \citep{carleton2012small_sample, twinkietown2024sample_sizes}. 
This concern about small sample variability is foundational in statistics, with the James-Stein estimator providing a classic example in baseball: even seemingly straightforward (batting) averages can be misleadingly variable when based on limited data \citep{efron1977stein}.

In this paper, we present efficient methods for detecting changepoints in both binary and continuous performance metrics at the individual player level. 
Although our empirical focus is on MLB data, the underlying problem, of sequential monitoring of metrics for changes, arises across many domains, including manufacturing process control \citep{page1954continuous, shang2017nonparametric}, finance \citep{banerjee2020change, kim2022unsupervised}, and climatology \citep{reeves2007review}.
While changepoint detection methods have been used in prior sports analytics research, they have primarily focused on identifying changes in aggregated statistics for the average player, team, or league across seasons \citep{chiou2016change, whalen2024empirical}. Recent work has extended changepoint detection to the individual athlete setting using Bayesian, model-based segmentation approaches for correlated performance metrics \citep{kim2021binary}. 

In contrast, our focus is on computationally scalable, real-time monitoring of individual players across an entire league. 
The scale of current sports data, characterized by hundreds of athletes and event-level metrics, creates a multiple testing problem where traditional full-sample methods often lead to over-flagging.
Rather than specifying a full generative model for player performance, which can be sensitive to misspecification and computationally intensive, we combine a likelihood-based scanning step with split-sample inference, yielding a novel framework that can be applied across a wide range of performance metrics and distributional settings.
We focus on pitch-level metrics, such as fastball velocity and plate discipline rates, as these are metrics that are commonly monitored by teams and are less sensitive to contextual aggregation than outcome-based statistics (such as batting average or earned run average), while still reflecting meaningful changes in observed performance.
Our approach draws on ideas from the changepoint detection literature \citep{hinkley1970inference, sen1975tests, killick2014changepoint, matteson2014nonparametric, shang2017nonparametric, jewell2022testing} but integrates them in a novel framework specifically designed for this application.

In particular, we incorporate split-sample inference with hypothesis tests that include a shift parameter, $\Delta$.
By including $\Delta$, we give users explicit control over the magnitude of changes considered meaningful.
This complements and extends existing R packages for changepoint detection \citep{lindeloev2024mcp_packages}, which are primarily designed for continuous, approximately normal data and do not typically incorporate split-sample inference. By explicitly supporting both continuous and binary outcomes within a split-sample framework, our approach provides a flexible alternative for applications where non-Gaussian data and false positive control are crucial considerations (e.g., tracking changes in plate discipline metrics). We demonstrate that this framework is operationally tractable for league-wide monitoring and provide open-source R functions that implement this framework efficiently at the player level.

We validate our method via simulation studies and on a constructed quasi-ground-truth dataset of pitchers who changed roles from reliever to starter during the 2023 or 2024 MLB seasons. We illustrate the broader utility of the method by detecting changes in fastball velocity and plate discipline metrics such as whiff and chase rate.

In practice, the proposed framework is designed for routine, in-season monitoring of player performance metrics. Changepoint detection can be run at regular intervals across selected metrics to flag observable shifts in performance across large numbers of players. These flags may reflect a range of underlying phenomena, including mechanical adjustments, fatigue, injury, or other changes affecting performance. By combining a fast likelihood-based scan with split-sample confirmation, the approach supports automated detection of potentially meaningful changes for subsequent manual review, providing a rigorous, scalable tool for monitoring player performance throughout the season.

The paper is organized as follows. Section~\ref{sec:data} describes the MLB dataset. 
Section~\ref{sec:change} introduces the changepoint detection algorithms. 
Section~\ref{sec:simulation} presents simulation studies evaluating Type I error control and power. We then apply the proposed methods to batter plate discipline metrics in Section~\ref{sec:change-rate} and to pitcher fastball velocity in Section~\ref{sec:change-continuous}. Validation using a quasi-ground-truth dataset of pitcher role transitions is presented in Section~\ref{sec:ground-truth}. Section~\ref{sec:discuss} concludes with a discussion of limitations and directions for future research.

\section{MLB Data}
\label{sec:data}
We consider pitch-level data from the 2023 and 2024 MLB seasons obtained from MLB Statcast, accessed via Baseball Savant. 
Pitches from all regular season games were included; postseason and spring training data were excluded.

Pitch data are ordered chronologically within each player by pitch timestamp. For batters, sequences are constructed separately for each plate discipline metric using only the subset of pitches relevant to that metric (e.g., pitches outside the strike zone for chase rate, and pitches swung at for whiff rate). For pitchers, sequences consist of all fastballs of a given pitch type (four-seam or sinker) thrown by that pitcher during the season.

For pitchers, we analyze fastball velocity at the pitch level, treating four-seam fastballs and sinkers as distinct pitch types. For batters, we analyze plate discipline metrics, which evaluate a player's ability to be selective with their swings, i.e., to swing at pitches they can hit well while taking pitches they cannot. For example, some plate discipline metrics evaluate a player’s ability to swing at pitches within the strike zone and refrain from swinging at pitches outside of the strike zone. The strike zone is defined as the area over home plate from approximately the batter’s knees to the midpoint of their torso, as measured by Statcast. 
In this paper, we focus on the following batter plate discipline metrics:

\begin{itemize}
    \item Out-of-zone swing: whether the batter swings at a pitch located outside the strike zone. Aggregating this metric across pitches (outside the strike zone) yields the out-of-zone swing percentage (O-Swing\%), also known as \textit{chase rate}.
    \item Swinging strike: whether the batter fails to make contact with a pitch that they swing at. Aggregating this metric across all pitches a batter swings at yields the swinging strike rate (SwStr\%), also known as \textit{whiff rate}.
\end{itemize}

\section{Changepoint detection}
\label{sec:change}

In this section we present a tractable changepoint detection algorithm for player performance data. 
Changepoint detection aims to identify the time point, if one exists, at which the statistical distribution generating the data changes.
Suppose we have a sequence of data $\{y_1, ..., y_n\}$.
The goal is to identify a point $t$, if it exists, such that 
$$y_1, ..., y_t \sim \mathcal{P} \text{ and } y_{t+1}, ..., y_n \sim \mathcal{Q}, \hspace{2mm} \mathcal{P} \neq \mathcal{Q}.$$
Here, we are primarily concerned with the problem of identifying changes in the mean of a player’s observed performance metric over time. For example, detecting when a batter's out-of-zone swing percentage decreases or when a pitcher's fastball velocity increases.

\subsection{Likelihood-Ratio Based Tests}
A common method for detecting changepoints uses a likelihood-based framework \citep{hinkley1970inference-1, hinkley1970inference}.
Consider a time-ordered sequence of data $y_{1:n} = (y_1, \dots, y_n)$.
Let $p(y \mid \theta)$ denote the probability density function of $y$ given parameter $\theta$, and let $\hat{\theta}$ denote the maximum likelihood estimate for the full dataset under the null hypothesis of no changepoint.

Under the alternative hypothesis, there exists a changepoint $t$ such that $y_{1:t}$ follow $p(y \mid \theta_0)$ and $y_{t+1:n}$ follow $p(y \mid \theta_1)$. 
Let $\hat{\theta}_0^{(t)}$ and $\hat{\theta}_1^{(t)}$ be the MLEs fit to each respective segment. 
Then the log-likelihood under the alternative is:
$$
\ell(t) = \log(p(y_{1:t} \mid \hat{\theta}_0^{(t)})) + \log(p(y_{t+1:n} \mid \hat{\theta}_1^{(t)} ))
$$

Under the null, the log-likelihood is

$$
\log(p(y_{1:n} \mid \hat{\theta})).
$$

Define

$$
\lambda_t = \ell(t) - log(p(y_{1:n} \mid \hat{\theta})).
$$

and take the test statistic as

$$
\lambda = \max_t \lambda_t.
$$

with the candidate changepoint

$$t' = \operatorname*{arg\,max}_t \lambda_t.$$

In many classical changepoint procedures, $\lambda$ is compared to a threshold $c$ to determine whether a changepoint is present. Choosing a threshold is challenging in practice \citep{shang2017nonparametric} and can lead to substantial overflagging in sports applications (see Sections~\ref{sec:simulation}-\ref{sec:change-continuous}).

In contrast, we do not threshold the likelihood-ratio statistic. Instead, we use the likelihood-ratio scan only to identify a candidate changepoint location $t'$, and perform all formal inference using a split-sample confirmation step described in the next section.

\subsection{Split-sample inference}
To help control the Type I error rate while maintaining computational tractability, we combine changepoint detection with split-sample inference \citep{cox1975note}. 
Split-sample inference divides the data into two parts: one used to identify a candidate changepoint location, and a second, held-out portion, used to test whether a change in the metric is present.

In our setting, we split the data into even and odd indices to ensure comparable sample sizes and to reduce dependence between the detection and confirmation steps.
The odd-indexed observations are used to calculate the candidate changepoint location $t'$, while the even-indexed observations are used to test the null hypothesis $\theta_0 = \theta_1$ (or, as described in Section~\ref{sec:change-continuous}, a shifted null $\theta_1 = \theta_0 + \Delta$ for a pre-specified $\Delta$).
This design avoids the need to select a global threshold for the likelihood-ratio scan; instead, the scan is used only to identify a candidate changepoint location, which is then evaluated using held-out data.

While even-odd splitting reduces dependence between the detection and confirmation stages, it does not eliminate serial correlation or other forms of dependence potentially present in pitch-level data. 
The confirmation step assumes that dependence across the split is sufficiently weak that the held-out observations provide approximately independent information about the presence of a change.
Thus, the confirmatory test should not be interpreted as exact under all realistic data-generating mechanisms, but rather as a practical mechanism for substantially reducing false positives relative to full-sample testing.
In Section~\ref{sec:simulation}, we evaluate the robustness of this approach under mild serial dependence.

The full single-changepoint detection procedure, incorporating likelihood-ratio scanning and split-sample inference, is presented in Algorithm~\ref{alg:cpt}.

\begin{algorithm}
\caption{Single Changepoint Detection Across Players}
\begin{algorithmic}[1]
    \State \textbf{Input:} $N$ (number of players), $\alpha$ (significance level), and the statistic of interest.
    \For{each player $i = 1, \ldots, N$}
        \State Obtain the time-ordered statistic sequence for player $i$.
        \State Split the sequence into two disjoint sets: odd-indexed and even-indexed timepoints.
        \State Let $T_{odd}$ be the set of odd-indexed timepoints.
        \For{each $t \in T_{odd}$}
            \State Compute the log-likelihood test statistic $\lambda_t$ (e.g., Equation~\ref{eq:lambda} for binary data and Equation~\ref{eq:lambda-norm} for normally distributed data).
        \EndFor
        \State Identify the candidate changepoint: $t^* = \arg\max_{t} \lambda_t$.
        \State Perform a hypothesis test at level $\alpha$ using the even-indexed timepoints.
        \If{the test is significant}
            \State Declare $t^*$ as the changepoint.
        \Else
            \State Declare no changepoint for player $i$.
        \EndIf
    \EndFor
\end{algorithmic}
\label{alg:cpt}
\end{algorithm}

\subsection{Multiple changepoint detection}

Algorithm~\ref{alg:cpt} details our method for detecting a single changepoint, which is a common use case in sports settings. In practice, teams often want to identify a changepoint, flag it, and continue monitoring for the next changepoint as new data arrive. Algorithm~\ref{alg:cpt} naturally extends to multiple changepoint detection via binary segmentation, which is helpful when scanning a long historical window for several performance shifts.

Binary segmentation works by repeatedly applying the single changepoint detection algorithm to successively smaller segments. Starting with the full series, we (1) run Algorithm~\ref{alg:cpt} on the current segment; (2) if a significant changepoint is found, we split the segment at that location; and (3) recursively analyze the left and right subsegments. We stop splitting a segment when no significant changepoint is detected (at level $\alpha$) or when either child would be shorter than a minimum length $m$, which ensures the test in Algorithm~\ref{alg:cpt} has adequate data on both sides of a candidate split. See Algorithm~\ref{alg:binseg}.

Because multiple tests are performed across segments, users may optionally control the overall false discovery rate (e.g., by adjusting $\alpha$ adaptively or applying a correction such as Bonferroni across detected splits).

In Section~\ref{sec:change-rate} and \ref{sec:change-continuous} we apply Algorithm~\ref{alg:binseg} to detect changes in batter plate discipline and pitcher fastball velocity respectively. 
We chose these metrics because they are more directly under the batter’s or pitcher’s control and less influenced by the opposing team or ballpark.
Code to implement all changepoint detection algorithms is available at  \url{https://github.com/akglazer/sports-changepoint}.

\begin{algorithm}
\caption{Multiple Changepoint Detection Across Players (Binary Segmentation)}
\begin{algorithmic}[1]
    \State \textbf{Input:} $N$ (number of players), $\alpha$ (significance level), $m$ (minimum segment length), and the statistic of interest.
    \For{each player $i = 1, \ldots, N$}
        \State Obtain the time-ordered statistic sequence $x_i$ of length $n_i$ for player $i$.
        \State Initialize the set of detected changepoints $\mathcal{C}_i \gets \varnothing$.
        \State Initialize a segment list $\mathcal{S} \gets \{(1, n_i)\}$.
        \While{$\mathcal{S}$ is not empty}
            \State Select a segment $(s,e)$ from $\mathcal{S}$.
            \If{$e - s + 1 < 2m$}
                \State \textbf{continue} \Comment{not enough length to split into two parts of size $\ge m$}
            \EndIf
            \State Apply Algorithm~\ref{alg:cpt} to the subsequence $x_i[s:e]$ to obtain a changepoint $\tilde{t}$ (local index in $[s,e]$).
            \If{no changepoint is returned at level $\alpha$}
                \State Do not split $(s,e)$ further.
            \Else
                \State Convert to global index: $t^* \gets s + \tilde{t} - 1$.
                \If{$t^* - s + 1 \ge m$} \State Add $(s, t^*)$ to $\mathcal{S}$. \EndIf
                \If{$e - t^* \ge m$} \State Add $(t^*+1, e)$ to $\mathcal{S}$. \EndIf
                \State $\mathcal{C}_i \gets \mathcal{C}_i \cup \{t^*\}$.
            \EndIf
        \EndWhile
        \State Output $\mathcal{C}_i$ for player $i$.
    \EndFor
\end{algorithmic}
\label{alg:binseg}
\end{algorithm}

\section{Simulation studies}
\label{sec:simulation}

In this section, we evaluate the performance of the proposed changepoint detection framework via simulation. The goals of this section are to (i) evaluate whether the split sample procedure adequately controls type I error rate; and (ii) examine the power of the algorithm to detect a changepoint when a true change is present.
We consider both binary and continuous data-generating processes, mirroring the plate discipline and fastball velocity applications presented in Sections~\ref{sec:change-rate} and~\ref{sec:change-continuous}.
Across all simulations, a changepoint is declared only if the full procedure described in Algorithm~\ref{alg:cpt} rejects the null hypothesis at level $\alpha = 0.05$ using the held-out sample.
Results are based on $1,000$ independent Monte Carlo replicates for each configuration.

\subsection{Type I error control}
\label{sec:sims-type1}

We first evaluate Type I error control under the null hypothesis of no changepoint.
A Type I error is recorded if the procedure declares a changepoint at level $\alpha = 0.05$ when no change is present.
We compare two variants of the pipeline:
(i) \textit{with sample splitting}, as in Algorithm~\ref{alg:cpt}, where the scan uses odd-indexed observations and the confirmation test uses even-indexed observations; and
(ii) \textit{without sample splitting}, where the same observations are used both to select the candidate changepoint and to perform the confirmation test.

\paragraph{Continuous null model.}
We generate $T$ observations from a stationary Gaussian model with constant mean,
\[
y_i \sim \mathcal{N}(\mu, \sigma^2), \quad i = 1,\dots,T,
\]
with $\mu = 93$ and $\sigma = 1$ and $T \in \{400, 800\}$.
For each replicate, we compute the likelihood-ratio scan statistic and candidate changepoint using the Gaussian likelihood, and apply the two-sample permutation test at level $\alpha$ to confirm the candidate changepoint.
Under sample splitting, the scan is performed on odd-indexed observations and the test on even-indexed observations; without sample splitting, both steps use the full sequence.

\paragraph{Binary null model.}
We generate binary data as independent Bernoulli trials,
\[
y_i \sim \text{Bernoulli}(p), \quad i = 1,\dots,T,
\]
with $p \in \{0.10, 0.40\}$ and $T \in \{400, 800\}$.
Candidate changepoints are obtained via the Bernoulli likelihood-ratio scan, and confirmation is performed using Fisher's Exact Test at level $\alpha$, with and without sample splitting as above.

Table~\ref{tab:type1} reports empirical rejection rates.
Across both data types, sample splitting yields empirical Type I error near the nominal level, whereas omitting sample splitting substantially inflates false positives.

\paragraph{Sensitivity to mild serial dependence.}
To assess robustness to short-range dependence, we simulate sequences under two types of dependence.

For continuous metrics, we generate autocorrelated sequences by simulating $AR(1)$ processes:

\[
y_t = \mu + \rho (y_{t-1} - \mu) + \varepsilon_t,
\qquad
\varepsilon_t \overset{iid}{\sim} \mathcal{N}(0,1),
\]
with $\mu = 93$ and $\rho = 0.1$. 

For binary sequences, we simulate from a first-order Markov model with constant marginal success probability $p$:
\[
\Pr(y_t = 1 \mid y_{t-1}) =
\begin{cases}
p + \rho, & y_{t-1} = 1, \\
p - \rho, & y_{t-1} = 0,
\end{cases}
\]
with $p=0.25$ and $\rho = 0.05$.
We repeat the Type I error experiments for $T=800$.

Results are reported in Table~\ref{tab:type1} and show that sample splitting substantially reduces false positives relative to the no-splitting baseline. Type I error is well controlled for binary sequences under mild dependence, while for continuous sequences sample splitting substantially mitigates, but does not fully eliminate, Type I error inflation under serial dependence.

\begin{table}[!ht]
\centering
\caption{Empirical Type I error rates under the null hypothesis of no changepoint, comparing the proposed sample-splitting procedure to a no-splitting baseline.}
\begin{tabular}{llll}
\hline
Metric & Setting & With splitting & Without splitting \\
\hline
Continuous & $T=400$ & 0.049 & 0.479 \\
Continuous & $T=800$ & 0.051 & 0.533 \\
Continuous & $T=800, AR(1)$ ($\rho = 0.1$) & 0.081 & 0.623 \\
Binary ($p=0.10$) & $T=400$ & 0.037 & 0.452 \\
Binary ($p=0.40$) & $T=400$ & 0.045 & 0.486 \\
Binary ($p=0.10$) & $T=800$ & 0.035 & 0.478 \\
Binary ($p=0.40$) & $T=800$ & 0.049 & 0.545 \\
Binary ($p=0.25$) & $T = 800$, Markov dep. $(\rho=0.05)$ & 0.051 & 0.674 \\
\hline
\end{tabular}
\label{tab:type1}
\end{table}

\subsection{Power}
Next we evaluate the power of the algorithm to detect a changepoint when a true change is present. Power comparisons are restricted to the sample-splitting procedure, as the no-splitting baseline does not maintain nominal Type I error under the null, as demonstrated in the previous section.

\paragraph{Continuous metrics.}
We generate sequences of length $T \in \{400,800\}$ from a Gaussian mean-shift model with a single changepoint at $c = \lfloor T/2 \rfloor$:
\[
y_t \sim 
\begin{cases}
\mathcal{N}(\mu,\sigma^2), & t \le c,\\
\mathcal{N}(\mu+\delta,\sigma^2), & t > c,
\end{cases}
\]
with $\mu=93$ and $\sigma=1$.
We consider effect sizes $\delta \in \{0.25, 0.5\}$.
The confirmation step uses a two-sample permutation test.

Table~\ref{tab:power-cont} shows that power increases rapidly with both the magnitude of the mean shift and the length of the observed sequence. For small effect sizes ($\delta = 0.25$), power is limited for shorter sequences but improves substantially as $T$ increases, reflecting the reduced variance in the split-sample confirmation step. For larger, practically meaningful shifts ($\delta = 0.5$), the procedure achieves high power even at $T=400$ and is nearly certain to detect the change by $T=800$. 

\begin{table}[!ht]
\centering
\caption{Empirical power for the continuous mean-shift model.}
\begin{tabular}{cccc}
\hline
$T$ & $\delta$ & Power \\
\hline
400 & 0.25 & 0.318 \\
400 & 0.5 & 0.88 \\
800 & 0.25 & 0.549 \\
800 & 0.5 & 0.988 \\
\hline
\end{tabular}
\label{tab:power-cont}
\end{table}

\paragraph{Binary metrics.}
Binary sequences of length $T \in \{800, 1600\}$ are generated from a Bernoulli rate-shift model with a changepoint at $c=\lfloor T/2 \rfloor$:
\[
y_t \sim 
\begin{cases}
\text{Bernoulli}(p_1), & t \le c,\\
\text{Bernoulli}(p_2), & t > c,
\end{cases}
\]
with $p_1=0.20 - \delta$ and $p_2 = 0.20 + \delta$ for $\delta \in \{0.03, 0.05, 0.08\}$.
Candidate changepoints are obtained via the Bernoulli likelihood-ratio scan, and confirmation uses Fisher's Exact Test at level $\alpha=0.05$.
Results are summarized in Table~\ref{tab:power-bin}.

As expected, power depends strongly on both the magnitude of the rate shift and the sequence length. Small shifts ($\delta = 0.03$) are difficult to detect, particularly for shorter sequences, reflecting the inherent noisiness of binary outcomes and the reduced effective sample size induced by split-sample inference. However, power increases rapidly with both $T$ and $\delta$: for moderate shifts ($\delta = 0.05$), power exceeds 80\% by $T=1600$, and for larger, practically meaningful shifts ($\delta = 0.08$), detection is very high even at $T=800$.

The reduced sensitivity for small shifts is an explicit trade-off of the split-sample design, which prioritizes Type I error control over aggressive detection. In applied MLB settings, this trade-off is appropriate: typical chase-rate sequences exceed 1{,}000 pitches per season after filtering, and changes of interest are often larger than a few percentage points. Thus, the $T=1600$ results are representative of practical use, where the method exhibits strong power while maintaining reliable false positive control.

\begin{table}[!ht]
\centering
\caption{Empirical power for the binary rate-shift model.}
\begin{tabular}{ccc}
\hline
$T$ & $\delta$ & Power \\
\hline
800 & 0.03 & 0.151  \\
800 & 0.05 & 0.479  \\
800 & 0.08 & 0.916  \\
1600 & 0.03 & 0.309  \\
1600 & 0.05 & 0.811  \\
1600 & 0.08 & 0.996  \\
\hline
\end{tabular}
\label{tab:power-bin}
\end{table}

\section{Detecting changes in batter plate discipline metrics}
\label{sec:change-rate}
In this section we apply Algorithm~\ref{alg:binseg} to detect changes in two batter plate discipline metrics: chase rate and whiff rate. 
First, consider \textit{chase rate}. 
For a particular batter, let $\{y_i\}_{i=1}^n$ be the set of all $n$ pitches outside of the strike zone with
\[
y_i =
\begin{cases}
1 & \text{if the batter swings at the $i$th pitch} \\
0 & \text{if the batter does not swing at the $i$th pitch}
\end{cases}
\]

We want to determine whether there exists a timepoint $t'$ such that 
$$\{y_i\}_{i = 1}^{t'} \sim \text{Bernoulli}(p_1) \text{ and } \{y_i\}_{i = t'+1}^n \sim \text{Bernoulli}(p_2) \text{ where } p_1 \neq p_2.$$
We define a batter's chase rate across $b-a+1$ timepoints spanning the $a$th (out of zone) pitch to the $b$th pitch to be:

$$
\Bar{y}_{a:b} := \frac{1}{b-a+1} \sum_{i = a}^b y_i
$$

Following the approach outlined in Section~\ref{sec:change}, for a timepoint $t$, we define
\begin{equation}
\label{eq:L}
    \ell(t) = \log( \Bar{y}_{1:t}^{\sum_{i = 1}^t y_i} ((1-\Bar{y}_{1:t})^{\sum_{i = 1}^t (1-y_i)}) + \log( \Bar{y}_{t+1:T}^{\sum_{i = t+1}^T y_i} ((1-\Bar{y}_{t+1:T})^{\sum_{i = t+1}^T (1-y_i)})
\end{equation}

and

\begin{equation}
\label{eq:lambda}
\lambda_t = \ell(t) - \log( \Bar{y}_{1:T}^{\sum_{i = 1}^T y_i} ((1-\Bar{y}_{1:T})^{\sum_{i = 1}^T (1-y_i)}).    
\end{equation}

The potential changepoint is $$t' = \operatorname*{arg\,max}_t \lambda_t.$$ 

As in algorithm~\ref{alg:cpt}, we use split-sample inference to determine whether $t'$ is a changepoint. 
In particular, we split $\{y_i\}_{i = 1}^T$ into odd and even pitches: $y^{odd} = \{y_i\}_{i \text{ mod } 2 = 1}$ and $y^{even} = \{y_i\}_{i \text{ mod } 2 = 0}$.
We use the odd pitches, $y^{odd}$, to calculate our potential changepoint $t'$.
Then, we use the even pitches $y^{even}$ to test the null hypothesis that the average chase rate after $t'$ is the same as the average chase rate before $t'$: $p_1 = p_2$.
Since the data are generated from a Bernoulli distribution, in the hypothesis testing step, we apply Fisher's Exact Test. 

We also apply the same procedure to batter's whiff rate. For whiff rate, let $\{y_i\}_{i=1}^n$ be the set of all $n$ pitches that a particular batter swings at. Then:

\[
y_i =
\begin{cases}
1 & \text{if the batter fails to make contact with the $i$th pitch} \\
0 & \text{if the batter makes contact with the $i$th pitch}
\end{cases}
\]

Using this definition of $y_i$, we apply the same procedure as previously described for chase rate.

\subsection{Plate discipline results}
Using data from the 2023 and 2024 MLB seasons, we ran Algorithm~\ref{alg:binseg} on all batters who faced at least 100 out-of-zone pitches and who had at least 100 swings for chase rate and whiff rate respectively. Table~\ref{tab:rate-cps} summarizes the results. Sample splitting reduces the number of batters flagged for changepoints by 89\% for chase rate and 88\% for whiff rate.

\begin{table}[ht]
\centering
\caption{Number of batters flagged for changepoints in chase rate and whiff rate using Fisher’s Exact Test with $\alpha = 0.05$, with and without sample splitting. In the no sample splitting condition, Fisher’s Exact Test is applied to the same data used to detect potential changepoints.}
\begin{tabular}{lccc}
\hline
\textbf{Metric} & \textbf{Batters Considered} & \textbf{\makecell{Changepoints \\ (Sample Splitting)}} & \textbf{\makecell{Changepoints \\ (No Sample Splitting)}} \\
\hline
Chase Rate & 647 & 50 (8\%) & 465 (72\%) \\
Whiff Rate & 687 & 58 (8\%) & 479 (70\%) \\
\hline
\end{tabular}
\label{tab:rate-cps}
\end{table}

The MLB regular season ran from March 30 to October 1 in 2023 and March 20 (for a two game series in South Korea between the Padres and Dodgers before the rest of the games began on March 28) to September 30 in 2024.
For chase rate and whiff rate respectively, 62\% and 64\% of the flagged changepoints occurred in May through August.

\begin{figure}
    \centering
    \includegraphics[width=0.7\linewidth]{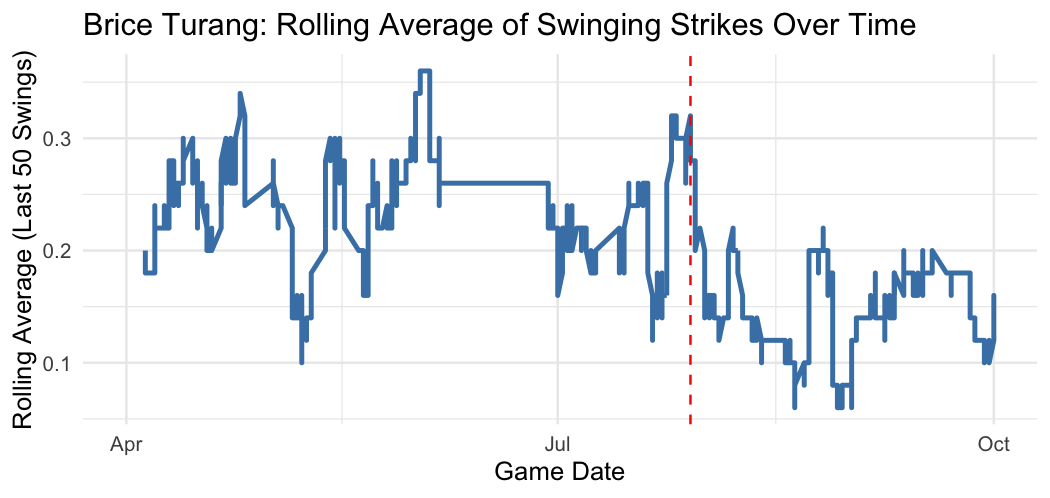}
    \caption{Brice Turang’s rolling average whiff rate, calculated over the last 50 swings, across the 2023 MLB season. The vertical red dashed line denotes the detected changepoint.}
    \label{fig:turang-whiff-rate}
\end{figure}

As an example, Figure~\ref{fig:turang-whiff-rate} plots a rolling average of Brice Turang's swinging strike rate for the 2023 season. 
The flagged changepoint is marked by the vertical dashed red line. 
While the changepoint detection algorithm operates directly on the binary pitch-level data, the rolling average is shown here as a useful way to visualize trends in a rate statistic.
Brice Turang saw an improvement to his whiff rate from the 2023 to 2024 season, with his whiff rate decreasing from 21.7\% to 13.8\%. Using our changepoint detection algorithm, we are able to flag this change at the end of July 2023.

\section{Detecting changes in pitcher fastball velocity}
\label{sec:change-continuous}

Next we consider the problem of detecting changes in a pitcher's primary fastball velocity. We assume that the velocity of a pitcher’s $i$th fastball follows a normal distribution, consistent with empirical pitch-level data \citep{statcastVeloDist}: $y_i \sim \mathcal{N}(\mu, \sigma^2)$ for some mean $\mu$ and variance $\sigma^2$.

Let 

$$
\hat{\mu}_{a:b} := \frac{1}{b-a+1} \sum_{i=a}^b y_i
$$

and

$$
\hat{\sigma}^2_{a:b} := \frac{1}{b-a+1} \sum_{i=a}^b (y_i - \hat{\mu}_{a:b})^2.
$$

Then, for a timepoint $t$, we can define

\begin{align}
    \ell(t) & = [-\frac{t}{2} \log(2\pi \hat{\sigma}^2_{1:t}) - \frac{t}{2}] +
    [-\frac{T-t}{2} \log(2\pi \hat{\sigma}^2_{t+1:T}) - \frac{T-t}{2}] \nonumber \\
    & = -\frac{t}{2} \log(2\pi \hat{\sigma}^2_{1:t}) -\frac{T-t}{2} \log(2\pi \hat{\sigma}^2_{t+1:T}) - \frac{T}{2}
\end{align}

and 

\begin{align}
\label{eq:lambda-norm}
    \lambda_t & = \ell(t) - [-\frac{T}{2} \log(2\pi \hat{\sigma}^2_{1:T}) - \frac{T}{2}] \nonumber \\
    & = -\frac{t}{2} \log(2\pi \hat{\sigma}^2_{1:t}) -\frac{T-t}{2} \log(2\pi \hat{\sigma}^2_{t+1:T}) + \frac{T}{2} \log(2\pi \hat{\sigma}^2_{1:T}).
\end{align}

The potential changepoint is $t' = \operatorname*{arg\,max}_t \lambda_t$. 

Again, we implement split sample inference according to Algorithm~\ref{alg:cpt}, with one important modification. Here, we test for a changepoint under a null hypothesis that allows for a location shift in the parameter of interest. Let  $\theta_0$ and $\theta_1$ denote the average metric values before and after the candidate changepoint, respectively. Under the null hypothesis, we assume 
$$\theta_1 = \theta_0 + \Delta,$$ 
where $\Delta$ is a pre-specified shift. We implement a two-sample permutation test that incorporates this shift directly. This approach allows for flexible hypothesis testing, e.g., one can choose to flag only changepoints where the post-change fastball velocity differs significantly by at least 1 mph (by setting $\Delta = 1$).

For metrics such as fastball velocity, where within-pitcher variability is low, even small changes can be detected with high statistical power when sufficient data are available, as demonstrated in our simulation studies in Section~\ref{sec:simulation}. However, small fluctuations are often not meaningful from a baseball perspective, especially with large sample sizes where statistical significance can outpace practical relevance. By incorporating a location shift $\Delta$ into the null hypothesis (e.g., testing whether the post-change average differs by more than 1 mph), we focus detection on changes that are practically significant. This modification reduces false positives driven by trivial variation and ensures that flagged changepoints correspond to deviations likely to reflect substantive changes, such as injury, mechanical adjustment, or sustained performance decline. For low-variance metrics like velocity, where the distinction between statistical and practical significance is especially pronounced, this approach improves interpretability and better aligns inference with domain expertise.

\subsection{Fastball velocity results}

We apply Algorithm~\ref{alg:binseg} using a two-sample permutation test with $\alpha = 0.05$ to detect changes in four seam fastball velocity for starting pitchers in the 2024 MLB season.
We define a starter as any pitcher that pitched more than 4 innings in an outing more than 10 times and entered in the 5th inning or later fewer than 5 times. We look at all pitchers with at least 100 four-seam fastball pitches.

Table~\ref{tab:norm-cps} summarizes the results. While the impact of sample splitting is not as drastic as with the plate discipline metrics considered in the previous section, sample splitting still reduces the number of flagged changepoints. Most notably, running the permutation test with shift parameter equal to one substantially reduces the number of detected changepoints.

\begin{table}[ht]
\centering
\caption{Number of starters ($N = 143$) in the 2024 MLB season flagged for changepoints in four-seam fastball velocity applying Algorithm~\ref{alg:binseg} using a two-sample permutation test with $\alpha = 0.05$, with and without sample splitting. In the no sample splitting condition, the permutation test is applied to the same data used to detect potential changepoints.}
\begin{tabular}{ccc}
\hline
\textbf{Threshold} & \textbf{Sample Splitting} & \textbf{Changepoints} \\
\hline
0 & Yes & 130 (91\%)  \\
0 & No & 142 (99\%) \\
1 & Yes & 22 (15\%)  \\
1 & No & 30 (21\%) \\
\hline
\end{tabular}
\label{tab:norm-cps}
\end{table}

Next, we consider starters across the 2023 and 2024 MLB seasons and run Algorithm~\ref{alg:binseg} with $\alpha = 0.05$ and two-sample permutation test with shift parameter equal to one.
Of the 174 starters considered, we flag 49 changepoints across 40 starters of which 21 (43\%) occur between September 1, 2023 and May 1, 2024 (i.e., during or near the offseason).

\section{Validation using known pitcher role transitions}
\label{sec:ground-truth}
One difficult aspect of changepoint detection in sports, and otherwise, is that it can be challenging or even impossible to acquire ground-truth data, because often we do not know if a change actually did occur.
In baseball, true changes in performance can occur for a multitude of reasons. For example, a pitcher's fastball velocity may
\begin{itemize}
    \item increase as he builds strength
    \item increase due to a biomechanical change he made in his approach
    \item decrease as he ages
    \item decrease due to an injury
    \item increase or decrease due to a change in role
\end{itemize}

It is difficult to construct a ground truth changepoint dataset for many of these types of changes because we do not know, for example, that a pitcher's fastball velocity will definitely decrease as he ages from 32 to 33. We often also do not know if a pitcher has made a biomechanical change (unless, for example, they tell the media or we have inside information). However, the last bullet point provides an avenue for constructing a quasi-ground-truth dataset to test our changepoint detection algorithm, as we know when a pitcher experiences a role change (e.g., from starter to reliever).
In this section we evaluate our algorithm's ability to detect known changes in pitcher statistics due to a role change using a curated ground-truth set of pitcher role changes.

We construct a dataset of pitchers, in the 2023 and 2024 MLB seasons, that switched from a relief to starting pitcher role to tune and test our changepoint detection algorithms. 
We consider fastball velocity, as it is well established that pitchers tend to lose velocity when transitioning from a relief role to a starting role (and gain velocity when moving from starter to reliever), since outings with a higher pitch volume typically require pitchers to reduce their intensity in order to remain effective over a longer duration compared to shorter relief appearances \citep{choiStarterReliever, palattella23}. 

Pitchers switch roles for a variety of reasons. For example, they might switch from pitching as a starter to as a reliever due to reduced stamina, struggles after the first or second time through the batting order, repeated injuries, roster construction issues, or significantly better performance in shorter outings. On the other hand, pitchers might switch from pitching as a reliever to as a starter due to development of another pitch, success in longer relief stints, a mechanical adjustment, or a team need for more starters.

We focus on building a quasi-groundtruth dataset around reliever to starter transitions as relievers frequently transition to a starting role due to improved performance and stamina, but face a drop in velocity after transitioning due to the endurance required to pitch through longer outings. On the other hand starters transitioning to relievers may switch roles due to a decline in performance and the boost to their velocity that comes from the role change might be washed out due to their performance decline.

To identify players that transition from a relief to starting role, we look at players that entered the game in the 5th inning or later (indicative of relief outings) more than 10 times and pitched more than 4 innings (indicative of a start) more than 10 times across the 2023 and 2024 seasons. We then filtered this list to only consider pitchers that increased their average number of innings pitched from 2023 to 2024 (i.e., moved from a reliever to starter).
We also excluded pitchers that transitioned roles more than once during this time period.
We confirmed and supplemented this list in consultation with news articles highlighting reliever to starter transitions \citep{relieverToStarter, relieverToStarter2}.

The players listed in Table~\ref{tab:rp-sp-transitions} were included in this ``ground-truth'' dataset.
The changepoint algorithm searched for changes in the velocity of their primary fastball (sinker for Jordan Hicks, Michael King and Jose Soriano; four-seam for the other pitchers) in the 2023 and 2024 seasons.
We implemented Algorithm~\ref{alg:binseg} with $\alpha = 0.05$ and increased the shift parameter of the permutation test until we were no longer able to detect fastball velocity changes for any player.
We were able to successfully flag the velocity changepoint for 91\% (10/11) of the pitchers. 
For 64\% (7/11) of the starter to reliever transitions, a fastball velocity change was flagged with shift parameter equal to 1 ($\Delta = 1$).
Results are displayed in Table~\ref{tab:rp-sp-transitions}.

\begin{table}[!ht]
    \centering
    \caption{Pitchers that transitioned from a relief to starting role at some point in the 2023 or 2024 MLB season. The ``Changepoint flagged?'' column indicates whether a changepoint was flagged in their exit velocity for their primary fastball (sinker for Jordan Hicks, Michael King and Jose Soriano; four-seam for all other players listed), and ``Max CP Threshold'' gives the maximum shift parameter ($\Delta$) that still flags a changepoint. Across all values of the shift parameter, Algorithm~\ref{alg:binseg} was implemented with $\alpha = 0.05$ and two-sample permutation tests.}
    \begin{tabular}{llll} \hline
         Pitcher name & MLB ID & Changepoint flagged? & Max CP Threshold  \\ \hline
         Jordan Hicks &  663855 & Yes & 5 \\ \hline
         Reynaldo Lopez &  625643 & Yes & 2 \\ \hline
         Ronel Blanco &  669854 & Yes & 0.5 \\ \hline
         Michael King & 650633 & Yes & 1 \\ \hline
         Zack Littell &  641793 & Yes & 1 \\ \hline
         Jose Soriano &  667755 & No & NA \\ \hline
         Garrett Crochet &  676979 & Yes & 0.5 \\ \hline
         Tyler Alexander &  641302 & Yes & 0.5 \\ \hline
         Andre Pallante &  669467 & Yes & 1 \\ \hline
         Cole Ragans &  666142 & Yes & 1 \\ \hline
         Sean Manaea &  640455 & Yes & 1 \\ \hline
    \end{tabular}
    \label{tab:rp-sp-transitions}
\end{table}

\begin{figure}[!ht]
    \centering
    \includegraphics[width=0.4\linewidth]{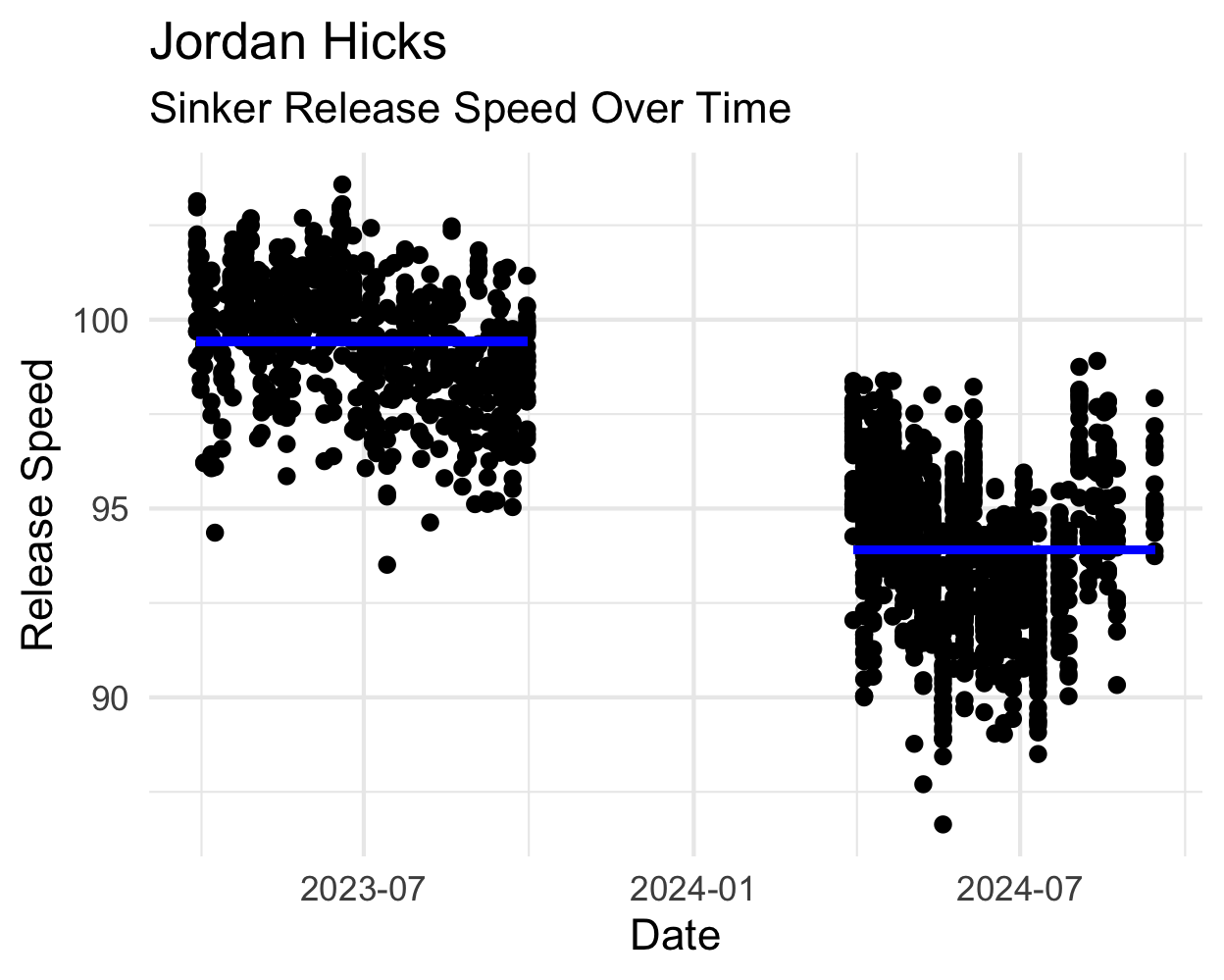}
    \includegraphics[width=0.4\linewidth]{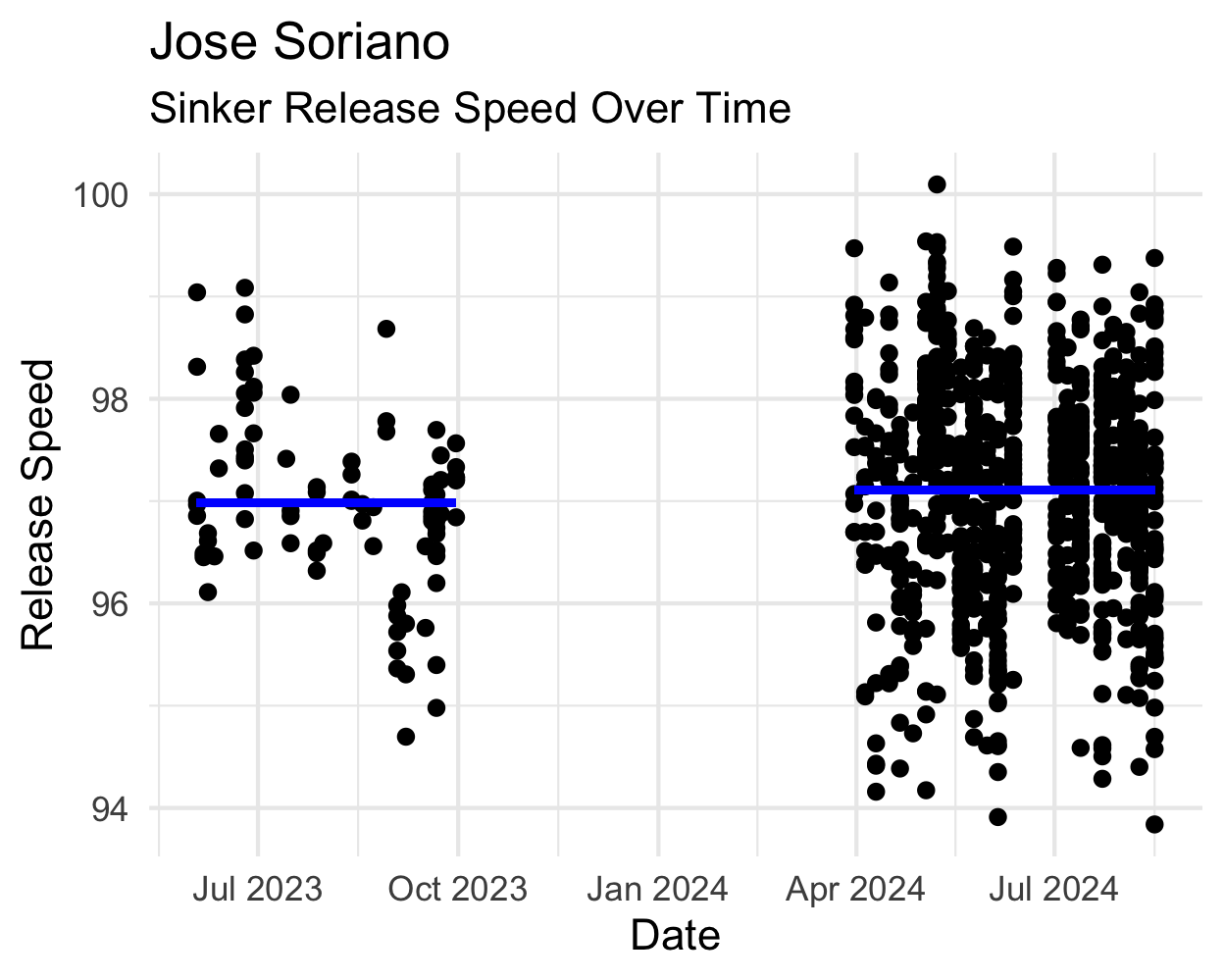}
    \caption{Jordan Hicks' and Jose Soriano's sinker velocity by pitch for MLB games in 2023 and 2024. Each plot features two horizontal blue lines. The leftmost and rightmost horizontal lines represent the average sinker velocity for each pitcher as a reliever and starter respectively. Hicks' average sinker velocity decreases by 5.5 mph after the transition. Soriano's average sinker velocity increases by 0.12 mph after the transition.}
    \label{fig:hicks-soriano}
\end{figure}

The largest change in fastball velocity was from pitcher Jordan Hicks. The only pitcher where a change in fastball velocity was not detected was Jose Soriano.
Figure~\ref{fig:hicks-soriano} plots fastball velocity by pitch for both Hicks and Soriano. Horizontal blue lines represent the mean sinker velocity before and after their role transition. While the average difference is quite apparent for Hicks (5.5 mph difference), it is negligible for Soriano (0.12 mph difference) making it clear why the algorithm did not flag a change in sinker velocity for Soriano.

\section{Discussion}
\label{sec:discuss}
In this paper we present tractable methods for detecting changepoints in binary and continuous player performance metrics. In particular, we propose a likelihood-based changepoint detection framework combined with split-sample inference, using either nonparametric tests or tests appropriate to the underlying data distribution. Hypothesis tests can additionally incorporate a shift parameter, allowing users to restrict attention to changes exceeding a pre-specified magnitude.

Through simulation studies, we show that the proposed procedure achieves reliable Type I error control under a range of data-generating settings and retains meaningful power to detect practically relevant changes. We then demonstrate the utility of the method on MLB data, applying it to batter plate discipline metrics and pitcher fastball velocity. Our approach flags velocity changes in 91\% of pitchers who transitioned from a relief to starting role during the 2023 or 2024 MLB seasons.
Furthermore, we find that for some metrics more than 60\% of detected changes occur in-season.

While this work draws on existing changepoint detection literature, methods are combined in a way specifically tailored to this setting.
In particular, the inclusion of split-sample inference with hypothesis tests that allow for a shift parameter allows users to better control the Type I error rate as well as have more control over the magnitude of changes detected.
Furthermore, we expand on existing R packages for changepoint detection \citep{killick2014changepoint, lindeloev2024mcp_packages} by providing functionality for binary sequences and split sample inference.
We provide an R implementation of all methods at \url{https://github.com/akglazer/sports-changepoint}.

We chose to focus on plate discipline metrics and fastball velocity because these metrics are commonly monitored by teams and are less directly influenced by contextual aggregation than outcome-based statistics such as batting average or earned run average. Nonetheless, these metrics are not fully context-free. For example, fastball velocity can exhibit some within-game fluctuations due to fatigue or environmental conditions. While our simulation studies suggest that the proposed split-sample procedure maintains Type I error control under mild dependence, future work could more explicitly model temporal dependence and contextual factors within the changepoint detection framework.

In this paper we primarily focus on detecting changes in means, however, the methods could easily be extended to look for changes in variance or other distributional characteristics. Furthermore, a nonparametric approach could be employed instead of the likelihood ratio-based method when the underlying data distribution is unknown. For example, one could use a divergence measure, as in \citet{matteson2014nonparametric}, instead of a likelihood ratio test statistic.

While permutation tests provide a flexible, distribution-free approach for confirming changepoints, they can be computationally intensive when applied repeatedly across many players and metrics.  In applications where distributional assumptions are reasonable, the confirmation step could instead use a parametric test, substantially reducing runtime while preserving the split-sample inference framework.

Although we focus on split sample inference as a way to reduce the false positive rate, alternative approaches could also be considered.
One such method, proposed by \citet{jewell2022testing}, evaluates the probability, under the null hypothesis of no true change in mean at a candidate changepoint, that the observed difference in means between segments on either side of a candidate changepoint could arise by chance. Specifically, it conditions on the set of changepoints and assesses whether the observed change is unusually large relative to all datasets that would yield the same changepoint configuration.

We could adapt this method to the binary setting by generating datasets $y' = \{ y'_i \}_{i = 1}^T$ such that $y'_i \sim \text{Bernoulli}(\Bar{y}_{1:T})$ and restricting attention to those in which a changepoint  $\lambda'$  falls within a window around the observed location, e.g., $\lambda' \in [\lambda - 50, \lambda + 50]$. We could then calculate $p = Pr\{ \Bar{y'}_{\lambda ' + 1:T} -  \Bar{y'}_{1:\lambda'} \geq \Bar{y}_{\lambda + 1:T} -  \Bar{y}_{1:\lambda} \}$, declaring  $\lambda$  a changepoint if  $p \leq \alpha$, where  $\alpha$  is the desired significance level.
A major challenge of this method, particularly at scale, is its computational cost, making it difficult to apply in real-time settings where analysts may wish to monitor many metrics across thousands of players daily. Nonetheless, future research could explore trade-offs between power and computation time, as well as potential algorithmic speed-ups or approximations that make this approach more feasible in practice.

While we focus on baseball applications in this paper, this work could easily be applied to other areas where it is of interest to detect changes in performance, e.g., manufacturing, healthcare or finance \citep{reeves2007review, shang2017nonparametric, banerjee2020change, kim2022unsupervised}.



\bibliography{bib}

@article{banerjee2020change,
  title={Change-point analysis in financial networks},
  author={Banerjee, Sayantan and Guhathakurta, Kousik},
  journal={Stat},
  volume={9},
  number={1},
  pages={e269},
  year={2020},
  publisher={Wiley Online Library}
}

@misc{carleton2012small_sample,
  author = {Russell A. Carleton},
  title = {Baseball Therapy: It’s a Small Sample Size After All},
  year = {2012},
  url = {https://www.baseballprospectus.com/news/article/17659/baseball-therapy-its-a-small-sample-size-after-all/},
  note = {Baseball Prospectus}
}

@incollection{chiou2016change,
  title={Change point analysis of top batting average},
  author={Chiou, Sy Han and Kang, Sangwook and Yan, Jun},
  booktitle={Extreme Value Modeling and Risk Analysis: Methods and Applications},
  pages={493--504},
  year={2016},
  publisher={CRC Press}
}

@misc{choiStarterReliever,
    title = {Breaking Down Baseball’s Early Velocity Surge},
    author = {Justin Choi},
    year = {2022},
    url = {https://blogs.fangraphs.com/breaking-down-baseballs-early-velocity-surge/}
}

@article{cox1975note,
  title={A note on data-splitting for the evaluation of significance levels},
  author={Cox, David R},
  journal={Biometrika},
  pages={441--444},
  year={1975},
  publisher={JSTOR}
}

@article{efron1977stein,
  title={Stein's paradox in statistics},
  author={Efron, Bradley and Morris, Carl},
  journal={Scientific American},
  volume={236},
  number={5},
  pages={119--127},
  year={1977},
  publisher={JSTOR}
}

@article{hinkley1970inference-1,
  title={Inference about the change-point in a sequence of random variables},
  author={Hinkley, David V},
journal = {Biometrika},
volume = {57},
issue = {1},
pages = {1-17},
  year={1970},
}

@article{hinkley1970inference,
  title={Inference about the change-point in a sequence of binomial variables},
  author={Hinkley, David V and Hinkley, Elizabeth A},
  journal={Biometrika},
  volume={57},
  number={3},
  pages={477--488},
  year={1970},
  publisher={Oxford University Press}
}

@article{jewell2022testing,
  title={Testing for a change in mean after changepoint detection},
  author={Jewell, Sean and Fearnhead, Paul and Witten, Daniela},
  journal={Journal of the Royal Statistical Society Series B: Statistical Methodology},
  volume={84},
  number={4},
  pages={1082--1104},
  year={2022},
  publisher={Oxford University Press}
}

@article{killick2014changepoint,
  title={changepoint: An R package for changepoint analysis},
  author={Killick, Rebecca and Eckley, Idris A},
  journal={Journal of statistical software},
  volume={58},
  pages={1--19},
  year={2014}
}

@article{kim2021binary,
  title={Binary segmentation procedures using the bivariate binomial distribution for detecting streakiness in sports data},
  author={Kim, Seong W and Shahin, Sabina and Ng, Hon Keung Tony and Kim, Jinheum},
  journal={Computational Statistics},
  volume={36},
  number={3},
  pages={1821--1843},
  year={2021},
  publisher={Springer}
}

@article{kim2022unsupervised,
  title={Unsupervised change point detection and trend prediction for financial time-series using a new cusum-based approach},
  author={Kim, Kyungwon and Park, Ji Hwan and Lee, Minhyuk and Song, Jae Wook},
  journal={IEEE Access},
  volume={10},
  pages={34690--34705},
  year={2022},
  publisher={IEEE}
}

@misc{lindeloev2024mcp_packages,
  author       = {Jonas Kristoffer Lindel{\o}v},
  title        = {An Overview of Change Point Packages in R},
  howpublished = {\url{https://lindeloev.github.io/mcp/articles/packages.html}},
  year         = {2024}
}

@article{matteson2014nonparametric,
  title={A nonparametric approach for multiple change point analysis of multivariate data},
  author={Matteson, David S and James, Nicholas A},
  journal={Journal of the American Statistical Association},
  volume={109},
  number={505},
  pages={334--345},
  year={2014},
  publisher={Taylor \& Francis}
}

@article{page1954continuous,
  title={Continuous inspection schemes},
  author={Page, Ewan S},
  journal={Biometrika},
  volume={41},
  number={1/2},
  pages={100--115},
  year={1954},
  publisher={JSTOR}
}

@misc{palattella23,
author = {Henry Palattella},
title = {'Flame on': How starters tackle a move to the 'pen},
year = {2023},
url = {https://www.mlb.com/news/how-starting-pitchers-transition-to-the-bullpen}
}

@article{reeves2007review,
  title={A review and comparison of changepoint detection techniques for climate data},
  author={Reeves, Jaxk and Chen, Jien and Wang, Xiaolan L and Lund, Robert and Lu, Qi Qi},
  journal={Journal of applied meteorology and climatology},
  volume={46},
  number={6},
  pages={900--915},
  year={2007}
}

@misc{relieverToStarter,
    author = {Brian Murphy},
  title = {7 Former Relievers Enjoying Life in the Rotation},
  year = {2024},
  url = {https://www.mlb.com/news/mlb-relievers-succeeding-as-starting-pitchers-in-2024}
}

@misc{relieverToStarter2,
  author = {Steve Adams},
  title = {Checking In On 2024’s Reliever-To-Rotation Experiments},
  year = {2024},
  url = {https://www.mlbtraderumors.com/2024/05/checking-in-on-2024s-reliever-to-rotation-experiments.html}
}

@article{shang2017nonparametric,
  title={Nonparametric change-point detection for profiles with binary data},
  author={Shang, Yanfen and Wand, Zhiqiong and He, Zhen and He, Shuguang},
  journal={Journal of Quality Technology},
  volume={49},
  number={2},
  pages={123--135},
  year={2017},
  publisher={Taylor \& Francis}
}

@article{sen1975tests,
  title={On tests for detecting change in mean},
  author={Sen, Ashish and Srivastava, Muni S},
  journal={The Annals of statistics},
  pages={98--108},
  year={1975},
  publisher={JSTOR}
}

@misc{statcastVeloDist,
  author       = {{MLB Advanced Media}},
  title        = {Statcast Pitch Velocity Distribution},
  howpublished = {\url{https://baseballsavant.mlb.com/visuals/statcast-pitch-distribution}},
  year         = 2025
}

@misc{twinkietown2024sample_sizes,
  author = {John Foley},
  title = {Analytics Fundamentals: When Do Stats Become Meaningful?},
  year = {2024},
  url = {https://www.twinkietown.com/2024/4/19/23030044/mlb-minnesota-twins-twinkie-town-analytics-fundamentals-sample-sizes-statistics-stabilization},
  note = {Twinkie Town}
}

@article{whalen2024empirical,
  title={Empirical Determination of Baseball Eras: Multivariate Changepoint Analysis in Major League Baseball},
  author={Whalen, Mena CR and Matthews, Gregory J and Mills, Brian M},
  journal={arXiv preprint arXiv:2407.01797},
  year={2024}
}

\end{document}